# Grain boundary migration in polycrystalline α-Fe

Zipeng Xu[1], Yu-Feng Shen[2], S. Kiana Naghibzadeh[3], Xiaoyao Peng[3], Vivekanand Muralikrishnan[1], S. Maddali[2], D. Menasche[2], Amanda R. Krause[1], Kaushik Dayal[3], Robert M. Suter[2], Gregory S. Rohrer[1]

[1]Department of Materials Science and Engineering, Carnegie Mellon University, Pittsburgh, PA 15213
[2]Department of Physics, Carnegie Mellon University, Pittsburgh, PA 15213
[3]Department of Civil and Environmental Engineering, Carnegie Mellon University, Pittsburgh, PA 15213

**Abstract**

High energy x-ray diffraction microscopy was used to image the microstructure of α-Fe before and after a 600 °C anneal. These data were used to determine the areas, curvatures, energies, and velocities of approximately 40,000 grain boundaries. The measured grain boundary properties depend on the five macroscopic grain boundary parameters. The velocities are not correlated with the product of the mean boundary curvature and grain boundary energy, usually assumed to be the driving force. Boundary migration is made up of area changes (lateral motion) and translation (normal motion) and both contribute to the total migration. Through the lateral motion component of the migration, low energy boundaries tend to expand in area while high energy boundaries shrink, reducing the average energy through grain boundary replacement. The driving force for this process is not related to curvature and might disrupt the expected curvature – velocity relationship.

## 1. Introduction

Grain boundary migration during annealing is an important process because of the role it plays in determining the microstructure [1]. When it occurs during grain growth, it reduces the grain boundary area and density of grain boundaries, and therefore, changes a range of macroscopic properties [2]. In the accepted theory for grain boundary migration, the grain boundary velocity (v) along the grain boundary normal direction is the product of the grain boundary mobility (M), energy (γ) and curvature (κ) [3]:

$$v = M \times \gamma \times \kappa \tag{1}$$

Experiments measuring the rate of grain boundary migration in Al bicrystals, reported by Gottstein and Shvindlerman [4], were consistent with Eq. 1. Later, Molodov et al. [5], discovered that the grain boundary velocity depended on lattice misorientation. Atomistic simulations reported by Upmanyu et al. [6], mimicking the geometry of the bicrystal experiments, were also consistent with Eq. 1.

The recent development of near field high energy x-ray diffraction microscopy (nf-HEDM) [7-9] and diffraction contrast tomography (DCT) [10-12] has made it possible to non-destructively measure the internal microstructure of polycrystals. Zhang et al. [13] used DCT to track grain growth in α-Fe and then used a phase field model to simulate [14] the experiment under the assumption that Eq. 1 is correct. However, the values of the grain boundary mobility that had to be assumed to reproduce the observations varied with time for the same boundary, raising the question of whether or not Eq. 1 is operative in polycrystals. Bhattacharya et al. [15] measured the velocity and curvature of more than 50,000 grain boundaries in a Ni polycrystal using nf-HEDM and found no correlation between grain boundary velocity and curvature, directly contradicting Eq. 1. A similar measurement of a $SrTiO_3$ polycrystal reported by Muralikrishnan et al. [16] led to the same conclusion. In $SrTiO_3$, 37 % of internal grain boundaries move in the direction opposite to that predicted by their curvature [16]. In addition, a 3D Monte Carlo Potts simulation of the experiment assuming isotropic grain boundary energy did not correctly predict the observed grain boundary motion [16].

Considering the experimental conditions, strain and defect related driving forces are not thought to be responsible for the recently reported and surprising observations outlined above [14-16], One possible explanation for these observations is that grain boundaries connected in a network are not always able to move as expected based on the curvature. While it is clear that isolated grain boundaries in bicrystals move towards their centers of curvature [17], a grain boundary in a polycrystalline network is unable to move without all the boundaries it is attached to also moving. It is plausible that because of the constraint of connectivity, not all of the grain boundary displacements can occur in the direction – and at the speed – predicted by Eq. 1. Evidence for this can be found in grain growth simulations [16, 18]. A second possibility is that the anisotropy of the grain boundary energy and mobility plays a role [19, 20]. The accepted theory for grain boundary migration assumes the properties are isotropic and that grain boundary curvature is uniform. These assumptions are known to be approximations and it is possible that accounting for anisotropy will lead to a much more complex correlation between curvature and velocity than predicted by the accepted theory.

The purpose of this paper is to describe how the rates for grain boundary migration in an $\alpha$-Fe polycrystal vary with boundary crystallography. We utilize the nf-HEDM measurements originally reported by Shen et al. [21], but here we compute the grain boundary populations, curvatures, velocities, and mobilities as a function of the grain boundary crystallographic parameters. The bcc structured Fe serves as contrast to similar data from Ni [15] and SrTiO$_3$ [16] which have the fcc and cubic perovskite crystal structures, respectively. Unlike the previous studies, we compare the velocities to the product of the curvature and anisotropic grain boundary energy. We also separately consider lateral grain boundary migration, reflected in grain boundary area changes, and normal grain boundary migration that occurs perpendicular to the grain boundary plane. Lateral migration is comparable to normal migration in magnitude and is correlated to the grain boundary energy and curvature. The results show that area increases are greatest for the lowest energy boundaries, supporting the idea that there is a driving force to replace relatively high energy grain boundaries with neighboring lower energy grain boundaries.

## 2. Methods

### 2.1. Experimental Data

The experiment and data collection were carried at part of Maddali's Ph.D. dissertation and the experimental details can be found in the dissertation [22] and a paper analyzing aspects of the data [21]. Here we summarize the parts of the work that are important for understanding the current results. The sample is bcc structured α-Fe that was electrolytically grown so that the microstructure never experienced the high temperature fcc phase. It was then rolled and recrystallized at 600 °C to homogenize the microstructure. Three-dimensional orientation maps of the microstructure were recorded in the initial state and after a 30 min anneal at 600 °C using nf-HEDM at the 1-ID beam line at Advanced Photon Source in Argonne National Laboratory. The program IceNine, a forward modeling method, was used to assign crystal orientations to each point in the analysis volume [23].

The open-source software Dream.3D [24] was then used to segment the microstructure into cube-shaped voxels with dimensions of $3 \times 3 \times 3$ μm$^3$. The minimum grain size was 8 voxels; smaller groups of voxels were absorbed into neighboring grains by grain dilation. The resulting initial and final state reconstructed microstructures have 10,927 and 9,224 grains within a volume of $1 \times 1 \times 0.195$ mm$^3$. In the current work, we used the entire microstructure to maximize the number of boundaries that could be analyzed, whereas in the work described in reference [21], the authors cropped the microstructure. The spherical equivalent diameter of grains with average volume in the initial and final microstructures are 29.8 µm and 31.6 µm, respectively.

The sample had a weak gamma fiber texture with (111) parallel to normal direction (3 MRD) and a weaker rotated cube texture. The gamma fiber texture increased to 3.5 MRD after annealing. Pole figures for the initial and final states are shown in Fig. S1. The distribution of grain boundary disorientations was similarly close to random, as illustrated in ref. [21].

2.2. Surface meshing and grain boundary curvature calculation

The voxelated grain interfaces after reconstruction were approximated by a smooth mesh using the Quick Surface Mesh filter in Dream.3D [24] and the grain boundary plane orientations were determined from this mesh. The meshed surface was smoothed by a Laplacian smoothing algorithm in Dream.3D, using 400 iterations and a weighting factor of 0.025 [24]. A detailed explanation of Laplacian smoothing can be found in reference [25]. Each triangle in the resulting mesh can then be associated with the grain identity and orientations on each side, centroid coordinates, area, and normal vector. Because this includes all five macroscopic grain boundary parameters (three for misorientation and two for the normal vector), it is possible to determine the grain boundary area, curvature, energy, and velocity distributions.

The Cubic-Order algorithm [26] of Dream.3D [24] was used to determine the principal curvatures for each meshed triangle. The algorithm calculates the Weingarten curvature matrix for each triangular mesh element. Two principal curvature values, $\kappa_1$ and $\kappa_2$, are the eigenvalues of the curvature matrix, and the triangle mean curvature value equals $(\kappa_1 + \kappa_2)/2$. Details about the accuracy of the curvature calculation can be found in reference [27]. It was assumed that the maximum observable curvature was determined by the spherical equivalent radius of a single voxel ($h_{voxel}$ = 0.55 μm$^{-1}$). Therefore, any triangle with a curvature greater than this upper limit was assumed to be non-physical and was excluded from the analysis. The excluded triangles are found almost exclusively at triple junctions. Under this criterion, the grain boundary mean curvature ($h_{GB}$) was determined by

$$h_{GB} = \frac{\sum_{i=1}^{n}(A_i * h_i)}{\sum_{i=1}^{n} A_i} \ (h_i \leq \max(h_{voxel})), \tag{2}$$

where $A_i$ and $h_i$ are the area and mean curvature of each triangle associated with the same grain boundary. In addition, boundaries are assumed to be indistinguishable when the left and right-hand crystals are exchanged (crystal exchange symmetry) so the mean curvature is positive definite for the purposes of computing distributions. In other words, for the curvature distribution, a grain boundary cannot be designated at convex or concave, because that characteristic depends on an arbitrary point of reference.

2.3. Grain boundary velocity calculation

During annealing, grains can either grow or shrink and the field of view can shift, so it is necessary to use several criteria to match grains in the initial and final states. Here, we used the process described previously [21, 28]. For grain $i$ in the initial state and grain $j$ in final state, the Confidence Index (CI) quantifying the likelihood that the i and j are the same grain is

$$CI_{ij} = p_1 * \Delta g_{ij} + p_2 * \Delta C_{ij} + p_3 * \Delta V_{ij}, \tag{3}$$

where $p_1$, $p_2$ and $p_3$ are three tunable parameters, $\Delta g_{ij}$, $\Delta C_{ij}$ and $\Delta V_{ij}$ are the disorientation angle, centroid difference, and volumetric difference between two grains, respectively. The assumption inherent in the method is that while none of these parameters will be identical in the initial and final state, the changes should be relatively small so that the two grains sharing the minimum CI are a match. Using these criteria, 8313 grains were matched leading to a grain matching efficiency of 90 %. Of the unmatched grains, 90 % were small grains near the minimum grain size defined during the Dream.3D segmentation process. The remaining unmatched grains were large grains that contacted the surface and were excluded based on that criterion. The unmatched grains are shown in Fig. S2.

Because the sample must be removed from the sample holder and then replaced after annealing, the position of the sample in the laboratory reference frame is likely shifted. To bring the initial state and final state reference frames into coincidence, we used the method described in reference [15]. The average disorientation between the voxels in the initial state and the voxels with the same position in the final state was calculated for 729 different translations of the form $\pm i\vec{x}$, $\pm j\vec{y}$ and $\pm k\vec{z}$, where i, j, and k are integers from 0 to 4 and the vectors are the unit translations on the lattice of voxels. The translation with the minimum average disorientation was assumed to be the correct alignment.

Using the aligned microstructures, the velocity of the boundary between grain i and grain j calculated as described in reference [15]

$$\upsilon_{GBij} = \left| \frac{V_{i \to j} + V_{j \to i}}{\bar{A}_{ij} * \Delta t} \right|, \tag{4}$$

where $V_{i \to j}$ and $V_{j \to i}$ are the volume of voxels that went from grain i to grain j and grain j to grain i, respectively. $\bar{A}_{ij}$ is the average grain boundary area between two time steps and $\Delta t$ is the annealing time. Note that the grain boundary velocity is the sum of the volume exchange between two neighboring grains instead of the difference to account for local motion of the boundary. Consider a grain boundary that has a shape like a sine wave and becomes flat after annealing. Although parts of the grain boundary migrate, the difference between the volumes exchanged is zero so no migration is detected. The sum of exchanged velocities, on the other hand, is positive reflecting the fact that parts of the boundary did migrate.

2.4. Grain boundary energy

To assign a grain boundary energy to each triangle, we used a piece-wise continuous grain boundary energy function previously defined for α- Fe [29] that specifies an energy for any given misorientation and grain boundary plane orientation. To specify the average energy of a boundary we use information from all of the triangles on that boundary

$$\gamma_{GB} = \frac{\sum_{i=1}^{n}(A_i * \gamma_i)}{\sum_{i=1}^{n} A_i} \ (h_i \leq \max(h_{voxel})), \tag{5}$$

where $A_i$ and $\gamma_i$ are the area and grain boundary energy (J/m²) of every triangle separating two grains. As before, triangles with absolute mean curvatures greater than maximum observable curvature were excluded in this analysis.

2.5. Grain boundary properties

At the end of the analysis, the data set comprised a list of grain boundary triangles, for which the relative area, curvature, energy, and velocity are specified. To determine a grain boundary property at any point in the five-parameter space, the program 3D_dist_graph [30] was used to analyze the list of triangles. This program, which performs the analysis as specified in reference [31], establishes apertures in the space of grain boundary disorientations and in the space of grain boundary plane orientations centered at the point of interest. It then finds all grain boundary triangles that fall within those apertures, determines the property mean value, and assigns that to the point in the five-parameter space. For all of the results here, the disorientation aperture was 5° and the grain boundary plane orientation aperture was 7°.

2.6. Simulation of microstructure evolution

The observed initial state microstructure was used as input for a threshold dynamics simulation of the evolution of the microstructure. The simulation method is described completely in reference [32] and the code used is available at [33]. During the simulation, a uniform and constant grain boundary energy of 1.0 was assumed for all grain boundaries. Therefore, an isotropic spherically symmetric Gaussian kernel was used as the convolution kernel in the threshold dynamics simulation. The simulation uses a voxelated structure, identical to the experimental data. Therefore, the simulation was started with the initial observed microstructure, which had an average grain size of 513 voxels per grain and terminated once the average grain size approximated that of the observed final microstructure, which was 610 voxels per grain. After 29 iterations with a normalized Gaussian kernel with step size of $10^{-6}$, the simulation was terminated when the average grain size reached 613 voxels per grain. The procedures used to analyze the final simulated microstructure were identical to those used to analyze the experimental microstructure.

## 3. Results

The grain boundary curvature and velocity distributions from more than 40,000 grain boundaries are illustrated in Figure 1. While the distributions are wide, more than 90 % of the grain boundaries had velocities less than 0.005 µm/s and curvature less than 0.05 µm$^{-1}$. To test for the expected linear correlation between grain boundary velocity and curvature, each boundary is plotted with these coordinates in Figure 2, where there is one point for each grain boundary. Fig. 2(a) shows all measured grain boundaries and Figure 2(b) shows a limited domain, to mitigate some of the overlap. There is no obvious linear correlation between grain boundary velocity and curvature. In fact, the distribution shows that there are more high velocity grain boundaries with small curvatures than with large curvatures.

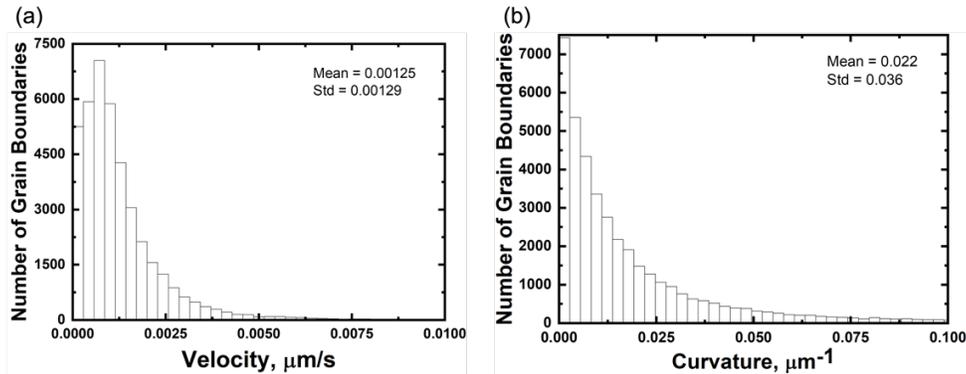

**Figure 1.** (a) Grain boundary velocity distribution, (b) Grain boundary curvature distribution. The mean and standard deviation (Std) are specified on each graph.

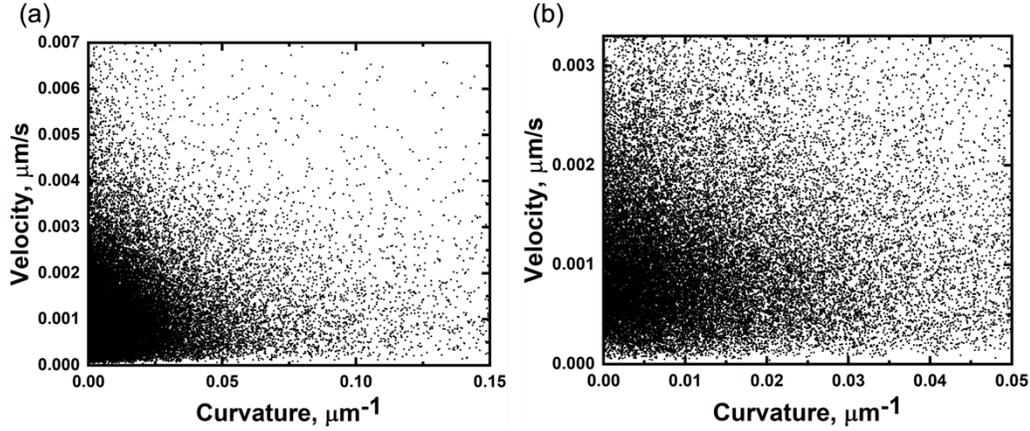

**Figure 2.** Scatter plots of grain boundary velocity and curvature. (a) All data. (b) Enlarged plot over a reduced domain that contains 90 % of the data points in (a).

To determine if averaged values of the velocity and curvature are correlated, the data were classified into discrete curvature groups, each with a width of 0.00275 $\mu m^{-1}$. The mean and standard deviation of the velocity was then determined for each group. We consider a domain including nearly 90 % of the data (35900 boundaries) with curvatures from 0 to 0.05 $\mu m^{-1}$, to ensure that each group contains a sufficient number of observations for averaging (at least 400 boundaries observation). In addition, velocities greater than three standard deviations from the mean were excluded, eliminating nearly 700 (2 %) of the boundaries. Figure 3(a) illustrates the mean values of the velocities in each group. The standard deviations at each point are large and are noted in the captions. A linear fit to the mean values yielded an $R^2$ value is only 0.14, indicating that there is no clear correlation between grain boundary velocity and curvature. We repeated this analysis employing a method to calculate the boundary curvature that operates only on the voxelated data and is therefore not subject to any possible bias that might be introduced by meshing or smoothing [16, 34]. The velocity-curvature correlation that results from this calculation is illustrated in Fig. S3 and supports the absence of correlation ($R^2 = 0.17$).

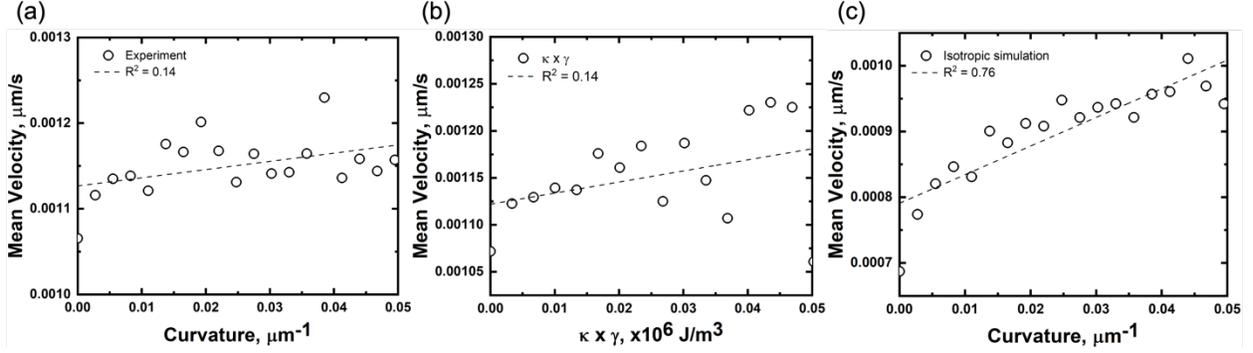

**Figure 3**. Mean grain boundary velocity versus curvature for (a) experimental data. (b) Mean grain boundary velocity versus the product of energy and curvature ($\kappa \times \gamma$). (c) Mean grain boundary velocity versus curvature for isotropic simulation. The dashed lines are linear fits that serve as guides to the eye. In each case, the standard deviations are large (not shown) but have roughly the same value at each point and average to $9.3 \times 10^{-4}$ $\mu m/s$, $9.2 \times 10^{-4}$ $\mu m/s$, and $7.2 \times 10^{-4}$ $\mu m/s$ in (a), (b), and (c), respectively.

The analysis in Fig. 2 and Fig. 3(a) ignores the fact that each boundary has a different energy, and this affects the driving force for migration. In its simplest form, the driving force is the product of the grain boundary energy and curvature. Using the grain boundary energy interpolation function [29], this was computed for each boundary and the correlation between the curvature energy product ($\kappa \times \gamma$) and the experimental grain boundary velocity is shown in Figure 3(b). Including the anisotropic grain boundary energy in the driving force does not improve the correlation with grain boundary velocity ($R^2 = 0.14$). Not only is the velocity of grain boundary migration not correlated to the curvature driving force, but we also find that some grain boundaries migrate away from their center of curvature (anti-curvature motion boundaries), as reported for $SrTiO_3$ [16]. The method we used to identify anti-curvature motion is illustrated in Fig. S4. For 40,280 measured boundaries, 45.1 % have anti-curvature motion.

The microstructure of the initial state was used as input for an isotropic threshold dynamics model used to simulate its time evolution. The correlation between the velocity and curvature in the simulation (Figure 3 (c)) has $R^2$ equal to 0.76, indicating a strong correlation. This correlation is consistent with Eq. 1. However, even though curvature and velocity are statistically correlated in the simulation, 38.0 % of the boundaries have anti-curvature motion. For grain boundaries with mean curvatures greater than 0.02 μm$^{-1}$, the fraction of boundaries with anti-curvature motion falls to 5.3 %. This confirms the result reported earlier using 3D Monte Carlo Potts simulations [16], that even with isotropic grain boundary properties, a boundary in a network cannot always migrate toward its center of curvature. It should be emphasized that the initial microstructure, the voxel representation, and the methods of analysis are identical for the experiment and simulation, so the difference in the outcome reflects a difference in the microstructure evolution, not in the analysis.

We have also examined the correlation between the velocity and crystallographic parameters in the α-Fe polycrystal. Fig. 4 shows the mean grain boundary curvature and velocity as a function of disorientation angle. Both quantities show smooth variations with the disorientation angle, and little correlation with each other. While the mean curvature differences are small, they correspond to a 36 % change in grain radius and simulations described in the supplemental section demonstrate that our measurements are sensitive to changes smaller than this (see Fig S5). Each point on the plot is the mean value of between $9 \times 10^4$ to $1 \times 10^6$ boundary triangles and the standard deviations are large in comparison to the absolute values. Therefore, while the mean values reflect the average of the population as a whole and show relatively smooth variations with disorientation, they are not good predictors of individual boundaries. The variation of the mean velocity with disorientation angle is similar to that predicted by bicrystal simulations [35], increasing from the low angle range, reaching a maximum near 20°, and then decreasing again for greater angles.

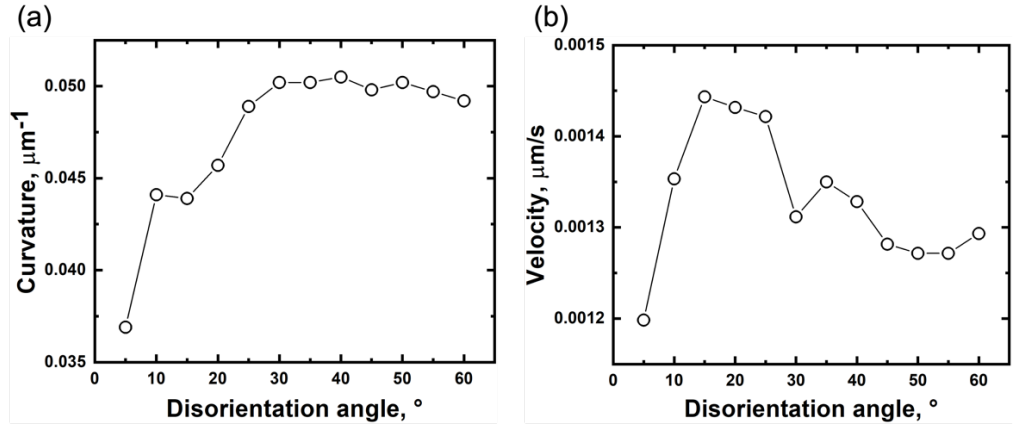

**Figure 4**. (a) Average grain boundary curvature and (b) average grain boundary velocity as a function of disorientation angle. The standard deviation at each point is about the same and for the curvature (velocity) it averages to 0.078 $\mu m^{-1}$ (0.001 $\mu m/s$).

Variations of the velocity with misorientation and grain boundary plane orientation are illustrated in Figure 5 for (a) twist and (a) symmetric tilt boundaries with misorientations around the [100] axis. Each mean value was determined from 10 to 120 distinct grain boundaries; the number of boundaries contributing to each point and the standard deviation at that point are provided in Fig. S6. For the symmetric tilt grain boundaries (STGBs), the Σ5 {013} boundary has a significantly larger velocity than the others, including the Σ5 {012} STGB. This is also the lowest energy Σ5 grain boundary [36]. The Σ5 twist boundary is one of the lowest velocity grain boundaries, indicating that there is significant variation in the velocity at fixed misorientation.

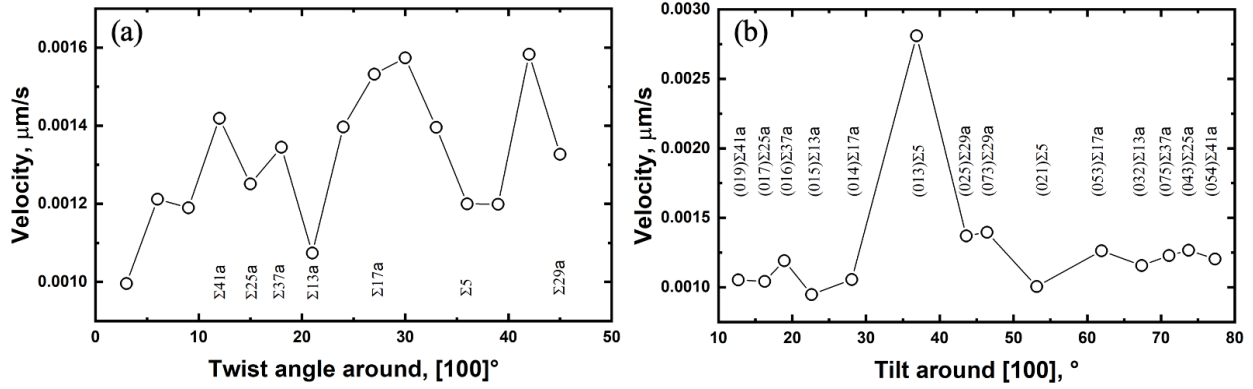

**Figure 5**. The variation of the grain boundary velocity with crystallographic parameters (a) Twist grain boundaries around the [100] axis sampled at a 3° interval. Coincidence site lattice (CSL) disorientations with $\Sigma \leq 49$, where $\Sigma$ is the inverse density of coincident lattice sites, are marked for reference. (b) STGBs with CSLs $\Sigma \leq 49$ around the [100] axis.

The variation of velocity with grain boundary plane orientation for the $\Sigma 5 = 36.9°/[100]$ disorientation is illustrated in Fig. 6(a). The grain boundary relative area, energy, curvature, and driving force ($\kappa \times \gamma$) are included for comparison in Fig. 6(b-e). These distributions were computed from 12,711 triangles from 173 distinct boundaries in the initial state. Two interesting findings arise from this comparison. First, the relative grain boundary area has an inverse correlation with curvature. However, the relative grain boundary area does not show the expected inverse correlation with the relative grain boundary energy, consistent with a previous study [29]. Second, there is a relatively strong correlation between energy and velocity for $\Sigma 5$ boundaries. For example, the {013} STGBs have the maximum velocity and the (100) twist boundaries are slower. These orientations also correspond to extrema of the grain boundary energy distribution. Both curvature and the curvature-energy product are uncorrelated with velocity. The correlation between energy and velocity does not arise in the driving force because the energy is nearly isotropic at this disorientation, so the driving force is dominated by the distribution of curvature.

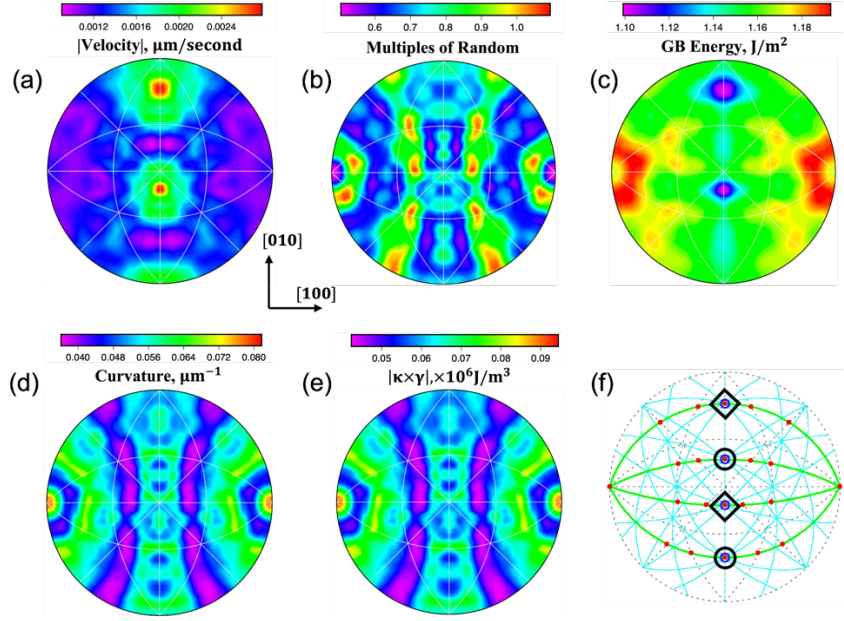

**Figure 6.** Grain boundary properties for 12,711 triangles from 173 Σ5 boundaries in the initial state, plotted on stereographic projections along [001]. (a) Grain boundary velocity distribution, (b) Grain boundary relative area distribution, (c) Grain boundary energy distribution, (d) Grain boundary curvature distribution, (e) Grain boundary curvature×energy distribution, (f) Stereographic projection generated in GBTool box. [37, 38] The orientations marked with diamonds and circles are {031} and {012} STGBs, respectively.

As a second example, we compare the variations of the grain boundary velocity and other properties as a function of grain boundary plane orientation at fixed misorientation (Σ3 = 60°/[111]) in Fig. 7. In this case, the distributions are calculated from 25,293 triangles from 180 distinct grain boundaries in the initial state. The velocity distribution has maxima at the $\{\bar{2}11\}$ STGB orientations, the (111) twist boundary orientation, and the asymmetric $(114)\|(0\bar{1}\bar{1})$ orientations. In addition, the $\{\bar{1}01\}$ tilt boundaries also have relatively large velocities. The asymmetric $(001)\|(1\bar{2}\bar{2})$ grain boundaries are the minima of the distribution. The $\{\bar{2}11\}$ STGBs, the (111) twist boundary, and the $\{\bar{1}01\}$ tilt boundaries are also local maxima and minima in the other properties. The twist boundary has the maximum energy and curvature, and a minimum population while the tilt boundaries have minimum energy and curvature and the maximum population, consistent with a previous studies of ferritic iron [27]. While the relative grain boundary area is inversely correlated with the energy and curvature, the velocity shows no clear correlation with these properties. Including the anisotropic grain boundary energy in the driving force (Fig. 7(e)) does not change the conclusion.

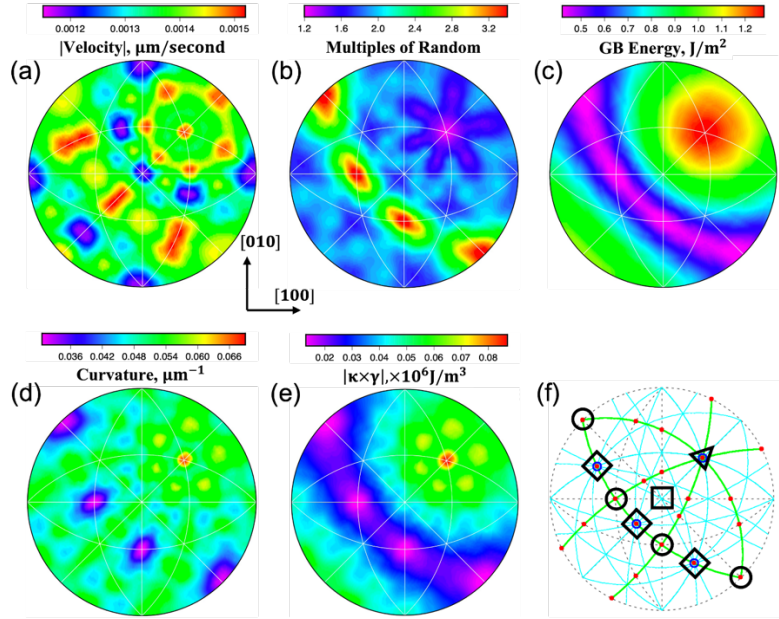

**Figure 7**. Grain boundary properties for 25,293 triangles from 180 distinct $\Sigma 3$ grain boundaries in the initial state, plotted on stereographic projections along [001] (Marked as square in (*f*)). (a) Grain boundary velocity distribution, (b) Grain boundary relative area distribution, (c) Grain boundary energy distribution, (d) Grain boundary curvature distribution, (e) Grain boundary curvature×energy distribution, (f) Stereographic projection generated in GBTool box. [37, 38] The orientations marked with circles, diamonds and triangle are $\{\bar{2}11\}$ STGBs, $\{\bar{1}01\}$ tilt boundaries and the [111] direction of the disorientation axis, respectively.

A key feature of grain growth is the reduction of grain boundary area, and this can occur either by translational motion normal to the grain boundary plane, or by area changes parallel to the boundary plane. We will refer to these two processes as normal and lateral migration, respectively. To explore the relative contributions of the two processes, we seek independent measures of each. For the normal migration, we use the cube root of the exchanged volume (the exchanged volume is the numerator of Eq. 4). To quantify the lateral component of the migration, we used the square root of the change in grain boundary area. When these two quantities are compared for each grain boundary, as in Fig. 8, it is clear that both processes can make a significant contribution to the total migration (the lateral component of the migration is greater than the normal component for 55 % of the boundaries).

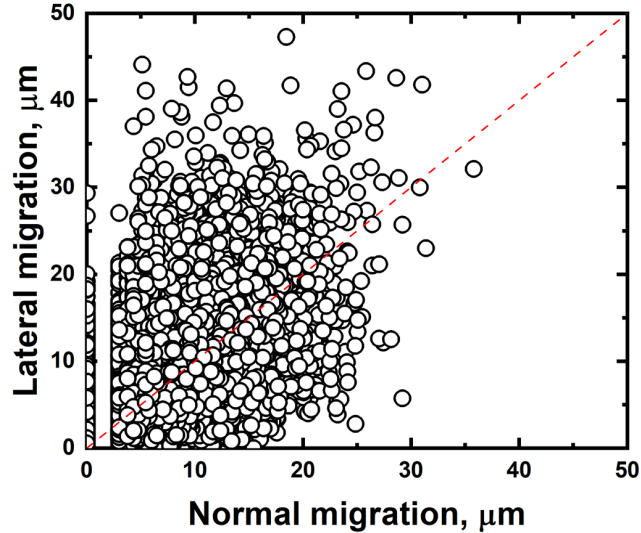

**Figure 8.** Comparison of lateral and normal migration for each boundary. The red dashed line has slope = 1. Because of the discrete sizes of the voxels, there are no data in the gap between 0 and 3 µm.

Considering the observation that the changes in area make a significant contribution to the total migration, the relationship between the grain boundary properties and the area change is examined in Fig. 9. When interpreting these data, it is important to note that they are biased against shrinking boundaries because we can only evaluate a velocity or area change for boundaries identified in both time steps. Boundaries that shrink below the resolution limit during annealing are necessarily excluded. Also, assuming the uncertainty for the average boundary is the area of one voxel, then the uncertainty of the area measurement is 3.5 %. Curvature, because of crystal exchange symmetry, is positive definite, so Fig. 9(a) shows the fractional area change as a function of absolute curvature; the data indicate that boundaries with larger curvatures have the greatest fractional change in area, and it is an increase in area. Fig. 9(b) shows that lower energy grain boundaries expand in area more than higher energy boundaries, a result consistent with the boundary replacement mechanism proposed to explain observations of Ni grain boundary migration [39]. There is also a clear correlation between the area changes and velocity (see Fig. 9(c)), with the fastest boundaries most likely to shrink and the slowest boundaries most likely to grow in area.

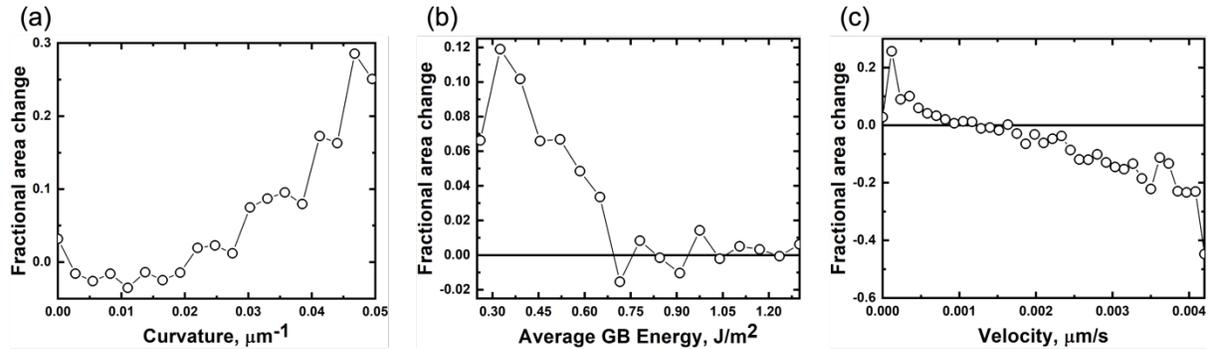

**Figure 9.** The correlation between grain boundary fractional area changes and different grain boundary properties. (a) Fractional area change $(\Sigma A_{final} - \Sigma A_{initial})/\Sigma A_{initial}$ as a function of curvature, (b) Fractional area change as a function of average grain boundary energy, (c) Fractional area change as a function of grain boundary velocity. The maximim estimated uncertainty of the fractional area change is 0.07.

An example of anti-curvature motion is illustrated in Fig. 10. The grain boundary between grain $i$ (left) and $j$ (right) is convex with the respect to grain $i$. The three-dimensional boundary shape after surface meshing is illustrated in Fig. S7 to show the boundary curvature is relatively uniform. The deviations from uniformity at the triple lines are typical and those falling above the threshold are excluded from the analysis. Assuming this grain boundary moves towards its center of curvature, it should move to the left. However, what we observe is that grain i increases its size and the convex grain boundary moves to right, which shows migration in a direction that is opposite to the one predicted by curvature. Note that instead of moving towards its center of curvature, the boundary migrated in such a way as to reduce its area and its energy by 15 %; as a result of migration, the boundary between i and j reduced its area from 609 μm² to 525 μm², leading to a reduction in its total energy from $6.9 \times 10^{-4}$ J to $5.9 \times 10^{-4}$ J. While this is a single anecdotal example, it exemplifies the idea that, because of the complex geometries of boundaries within polycrystals, boundaries can migrate to reduce their energy without moving towards their center of curvature.

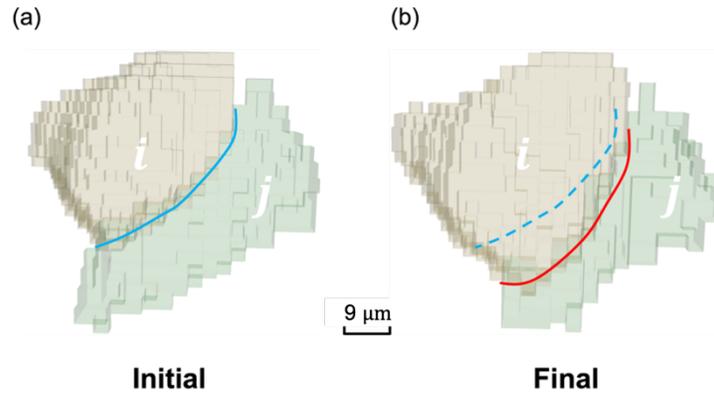

**Figure 10**. An example of a grain boundary that migrates away from its center of curvature. (a) Grain $i$ (left) and $j$ (right) in the initial state, the solid blue line represents the grain boundary between those two grains; (b) Grain $i$ (left) and $j$ (right) in the final state, the dashed blue line and solid red line represent the initial grain boundary and final grain boundary, respectively.

## 4. Discussion

Previous studies of grain boundary migration in Fe-3wt% Si bicrystals illustrated that the reduced mobility depended on temperature and misorientation [40, 41]. Using data from reference [40], we were able to extrapolate reduced mobilities to the lower temperature range of our experiment (about 400 °C lower than in reference [40]). For a range of [100] misorientations, the reduced mobilities of the bicrystals ranged from $1 \times 10^{-17}$ to $1 \times 10^{-12}$ m²/s. The reduced mobility for our data, measured from the velocity-to-curvature ratio, varied over a smaller range and the average ($5 \times 10^{-14}$ m²/s) was within the range observed for the bicrystals. While one should not necessarily expect data from an iron alloy, extrapolated to a lower temperature, to compare well with data from pure Fe, it is interesting that they at least fall in the same albeit wide range. Similarly, the consistency between the measurements of the velocity as a function of disorientation (see Fig. 4(b)) with molecular dynamics simulations [35] indicates that the current measurement do not conflict with earlier observations.

The results presented here are the third instance of the absence of a correlation between curvature and grain boundary migration velocity in polycrystals. While the initial observations in Ni [15] and SrTiO$_3$ [16] were unexpected, the consistency of the results in two metals with different crystal structures and a ceramic suggests that this is a general phenomenon. One point not addressed by the previous work is whether or not accounting for the effect of anisotropic grain boundary energy on the driving force ($\kappa \times \gamma$ in Eq. 1) would improve the correlation. As illustrated in Fig. 3(b), the velocity remains uncorrelated with the curvature driving force. We note that the grain boundary energies used here are an approximation of the true energies. For example, at 600 °C, the total anisotropy should be smaller than given by the interpolation function, which is fit to energies derived from 0 K embedded atom potential molecular dynamics simulations [29, 36]. However, the shape of this function is consistent with experimental studies of the grain boundary energy anisotropy [42] and the grain boundary character distribution of ferrite [43] and is expected to improve the correlation between the curvature driving force and velocity if that were the key factor missing in the analysis.

The results shown in Figs. 3, 4, and 5 (and supplemental figures) are averages over many boundaries or boundary triangles and it must be noted that the standard deviations of instances contributing to each of the data points is, as noted earlier, comparable to or larger than the trends in the averages. Thus, the averages are not good predictors of the behavior of individual instances. The observed trends in the averages, however, with noise levels well below the distribution widths, appear to be reliable and, given similar observations for many different quantities, should be repeatable in different samples and materials [13-15, 21, 28]. Any predictive theory of grain growth will have to reproduce both the average trends and the broad distributions to reproduce the observed microstructure evolution. The cause of these distributions might be related to variations of local neighborhoods surrounding the different instances (which raises the question of how large a neighborhood would be required to reproduce the average behavior). It is also possible that extrinsic and unobserved structural variations may also be influential, such as local variations in the three microscopic degrees of freedom of the boundary [44] or the availability of defects necessary for boundary migration, such as disconnections [45].

One important observation in this work is that lateral motion of the boundaries (area change) is as large as the normal displacement. This is highlighted by the data in Fig. 8. If we look simultaneously at the data in Fig. 9(a) and (c), we see that boundaries that increased (decreased) their area had high (low) curvature but low (high) normal velocity. This is a direct contradiction of Eq. 1. The boundaries that had a positive area change (velocity less than 0.0012 μm/s) make up 63.3 % of all observations. Therefore, this lateral motion, which makes up a significant fraction of the migration, is inconsistent with Eq. 1 and is the most likely reason that the expected correlation between curvature and velocity is disrupted. The driving force for these expansions and contractions is related to the grain boundary energy anisotropy and the energy differences between a certain boundary and the two it meets at a triple line, which is unrelated to curvature.

An example of this lateral motion is illustrated in Fig. 11. All of the boundaries in the image were colored by their energy and the grey boundary, with a relatively low energy of 0.58 J/m$^2$ (this is a low angle GB with a misorientation of 8.3° around [0.40 $\overline{0.48}$ 0.78], an axis that is about 4.8° from the [1$\overline{1}$2] direction), is labeled I. There are three neighboring boundaries with higher energies, labeled II, III, and IV. After annealing, boundary I expanded in area, almost replacing boundaries II, III, and IV. Although the low energy boundary increased its area, it annihilated area on the neighboring high energy boundaries, which reduces the total energy. We recently reported the same mechanism in Ni and referred to it as grain boundary replacement [39]. Using this mechanism, the grain boundary network can lower its energy by replacing high energy grain boundaries with lower energy grain boundaries, even if there is no change in the total area. In this case, the total grain boundary area decreased during annealing and the average grain boundary energy decreased from 1.094 J/m$^2$ in the initial state to 1.091 J/m$^2$ in the final state. While this decrease (0.3 %) is smaller than the energy change reported for Ni (2.8 %) [39], the Ni sample went through five anneal cycles, leading to more extensive changes in the microstructure. Because lateral migration is driven not by curvature but by the energy gained when high energy grain boundaries are replaced by lower energy grain boundaries, this is the most likely process that disrupts the expected curvature-velocity correlation.

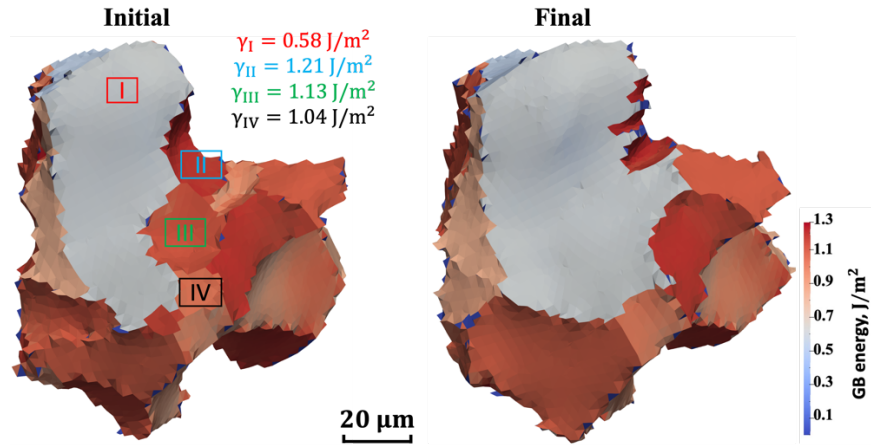

**Figure 6.** Example of low energy grain boundaries replacing neighboring grain boundaries with greater energy. Grain boundaries are represented by triangular meshes and colored based on their grain boundary energy.

As noted in the introduction, the anisotropy of the grain boundary energy and mobility might play a role in disrupting the influence of curvature on grain boundary migration. Simulations of microstructure evolution have shown that the grain boundary energy anisotropy is more influential than the mobility [46-48]. This strong influence of the grain boundary energy anisotropy likely arises from the effects that it has on triple junction geometry. The persistence and expansion of lower energy grain boundaries during grain growth has been previously reported based on two-dimensional observations (although that conclusion was reached using the assumption that boundaries migrated toward their centers of curvature) [49]. Consistent with these observations, recent 3D simulations using anisotropic energies found that low energy boundaries increase as a fraction of the population during grain growth [50]. Two other recent simulations of microstructure evolution have reported the motion of boundaries opposite to their curvature, also consistent with the experimental observations [16, 18]. In fact, even for the case of an isotropic simulation [16], anti-curvature motion was found, although the velocities of the majority of the boundaries were still correlated with curvature in that simulation, as also found here. This is evidence that the geometry of the interconnected network of boundaries (even in the absence of energy anisotropy) can lead to anti-curvature migration, but this alone is not sufficient to break the statistical correlation between curvature and velocity. The results presented here support the idea that grain boundary energy anisotropy and the grain boundary replacement mechanism are important in determining the direction and speed of grain boundary migration.

## 5. Conclusion

Based on measurements of the migration velocities and curvatures of approximately 40,000 grain boundaries in $\alpha$-Fe before and after a 30 min anneal at 600 °C, we find that both properties vary with the grain boundary disorientation and grain boundary plane orientation. For example, grain boundaries with disorientations less than 15° are relatively fast compared to higher angle grain boundaries and the $\Sigma 5$ {013} STGB is more than twice as fast as the $\Sigma 5$ twist boundary.

Consistent with two recent studies of polycrystals [15, 16], the boundary migration rates are uncorrelated with the mean curvature. The absence of a correlation persists when the anisotropy of the grain boundary energy is included in the driving force. Grain boundary area changes lead to increases in the areas of low energy grain boundaries and a decrease in the areas of higher energy boundaries, reducing the average grain boundary energy by the boundary replacement mechanism. The replacement of high energy grain boundaries with lower energy boundaries by lateral migration is the most likely process that disrupts the normal motion of a GB towards its center of curvature.

**Acknowledgments**

This work was supported by the National Science Foundation under DMREF Grant No. 2118945. The Advanced Photon Source is a US Department of Energy (DOE) Office of Science User Facility operated for the DOE Office of Science by Argonne National Laboratory under contract no. DE-AC02-06CH11357. The authors thank David Kinderlehrer for helpful discussions.

Supplemental information for :
# Grain boundary migration in polycrystalline α-Fe

Zipeng Xu[1], Yu-Feng Shen[2], S. Kiana Naghibzadeh[3], Xiaoyao Peng[3], Vivekanand Muralikrishnan[1], S. Maddali[2], D. Menasche[2], Amanda R. Krause[1], Kaushik Dayal[3], Robert M. Suter[2], Gregory S. Rohrer[1]

[1]Department of Materials Science and Engineering, Carnegie Mellon University, Pittsburgh, PA 15213
[2]Department of Physics, Carnegie Mellon University, Pittsburgh, PA 15213
[3]Department of Civil and Environmental Engineering, Carnegie Mellon University, Pittsburgh, PA 15213

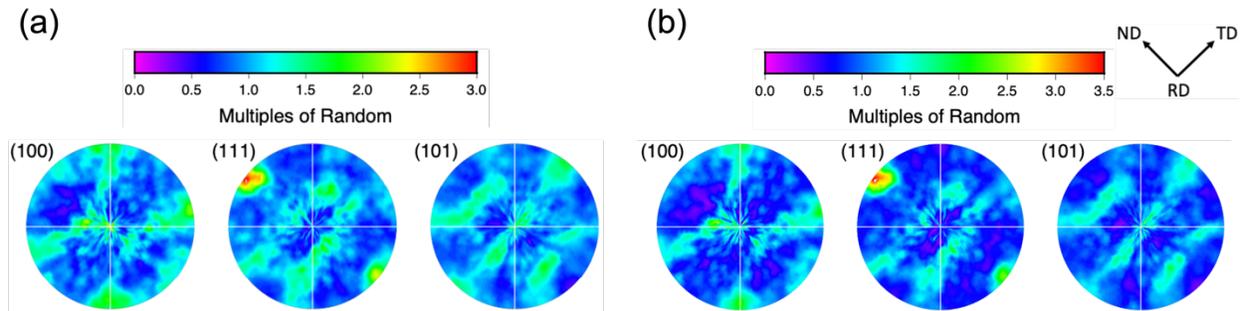

**Figure S1.** Pole figure of the $\alpha$-Fe sample computed from the 3D HEDM data. (a) The (100), (111), and (101) pole figures in initial state and (b) final state. The rolling direction (RD) is pointed out of the plane and the normal direction (ND) and transverse direction (TD) are shown by the arrows. The most significant texture is the preference for the (111) direction along the sample normal direction.

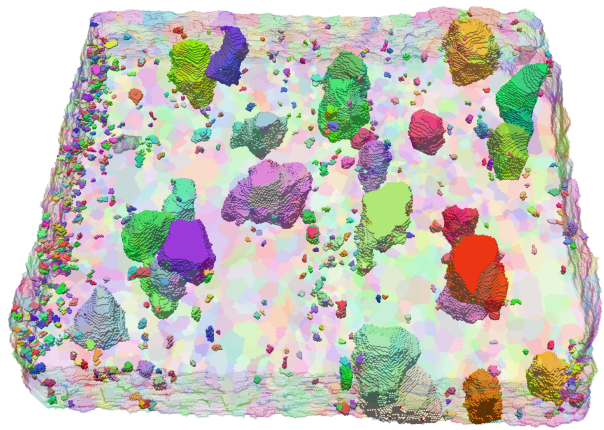

**Figure S2.** Visualization of unmatched grains in final state in $\alpha$-Fe. The matched grain are rendered transparent. The dimensions of the displayed volume are $1 \times 1 \times 0.195 \ mm^3$.

Figure S2 shows the unmatched grains in final state after the grain matching process. 90 % of the unmatched grains were small grains near the minimum grain size defined during the Dream.3D segmentation process and the rest were large grains that contacted the surface.

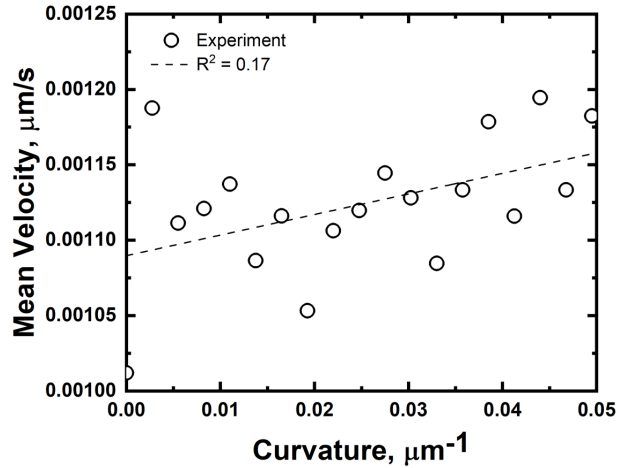

**Figure S3.** Mean grain boundary velocity versus curvature for experimental data. The curvature was calculated using the innie-outie formula, described in [1]. The average standard deviation at each point is $8.1 \times 10^{-4}$ $\mu m/s$.

Fig. S3 illustrates the mean velocity and curvature, using the same bin width as Fig. 3(a), but with a different technique for calculating the curvature. In this case, the poor $R^2$ value (0.17) suggests that the absence of a velocity curvature correlation does not depend on the calculation method.

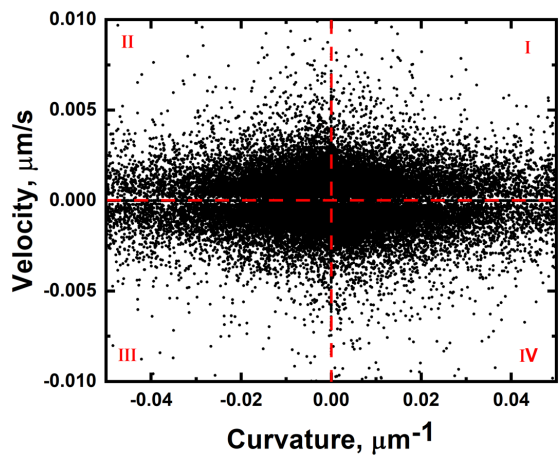

**Figure S4.** Signed velocity and curvature results, using the following conventions. For each boundary, if the motion reduces (increases) the number of voxels in the grain, the velocity is assigned to be negative (positive). The sign of the curvature for boundaries that are concave (convex) with respect to the grain center is positive (negative).

Figure S4 illustrates the conditions used to differentiate anti-curvature motion from motion consistent with the sign of the curvature. Each boundary appears twice in the plot, once when viewed from one grain and once when viewed from the other, so that the +/+ (I) and -/- (III) quadrants contain the identical data. The boundaries that fall in the regions I and III are the anti-curvature motion boundaries.

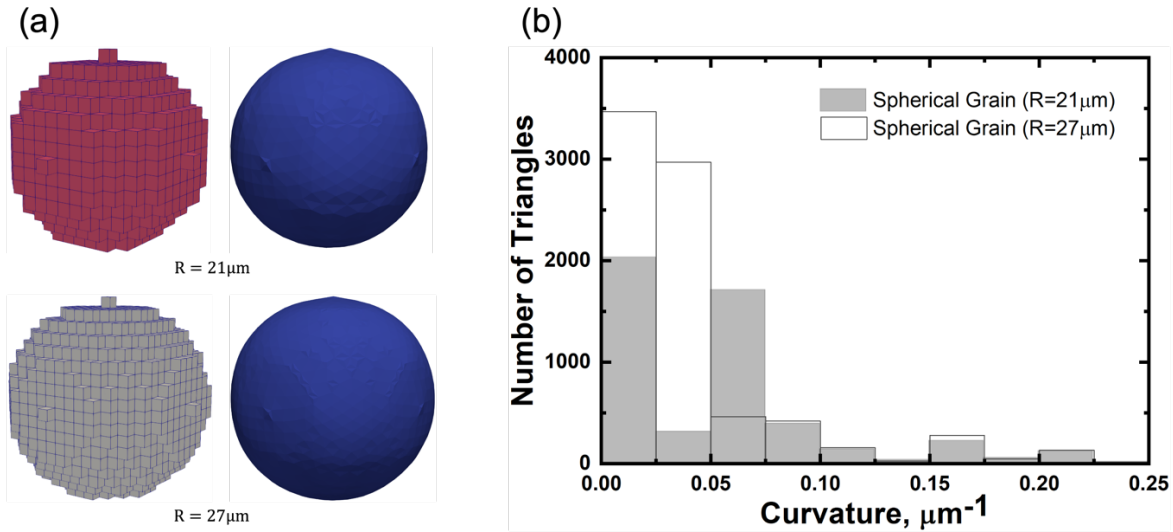

**Figure S5.** Curvature distribution of dummy spherical grains. (a) Dummy spherical grains. Upper: voxelated grain with $R = 21\ \mu m$ (left), surface meshed grain with $R = 21\ \mu m$ (right). Lower: voxelated grain with $R = 27\ \mu m$ (left), surface meshed grain with $R = 27\ \mu m$ (right). (b) Curvature distribution for the two spherical grains.

Fig. S5 shows curvature distributions for two simulated grains with different sizes. Spheres of these dimensions have ideal curvatures of 0.0476 and 0.0370. The mean curvature for the meshed spheres are 0.0442 and 0.0344, differing from the ideal by about 7 %. From this we conclude that the curvature differences in the Figure 4(a) can be measured by our method.

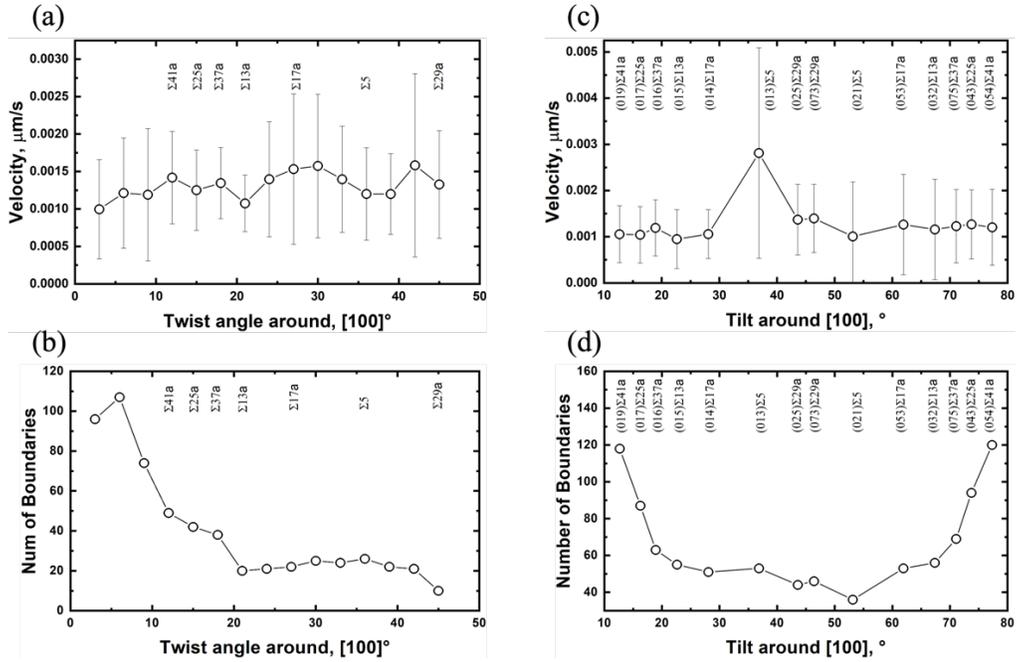

**Figure S6.** (a) and (c): data in Figure 5, with bars denoted ± one standard deviation. (b) and (d) the number of grain boundaries averaged to determine the mean value for each measurement.

Figure S6 provides details for the data in Fig. 5, illustrating the quantity of data and the estimated uncertainties.

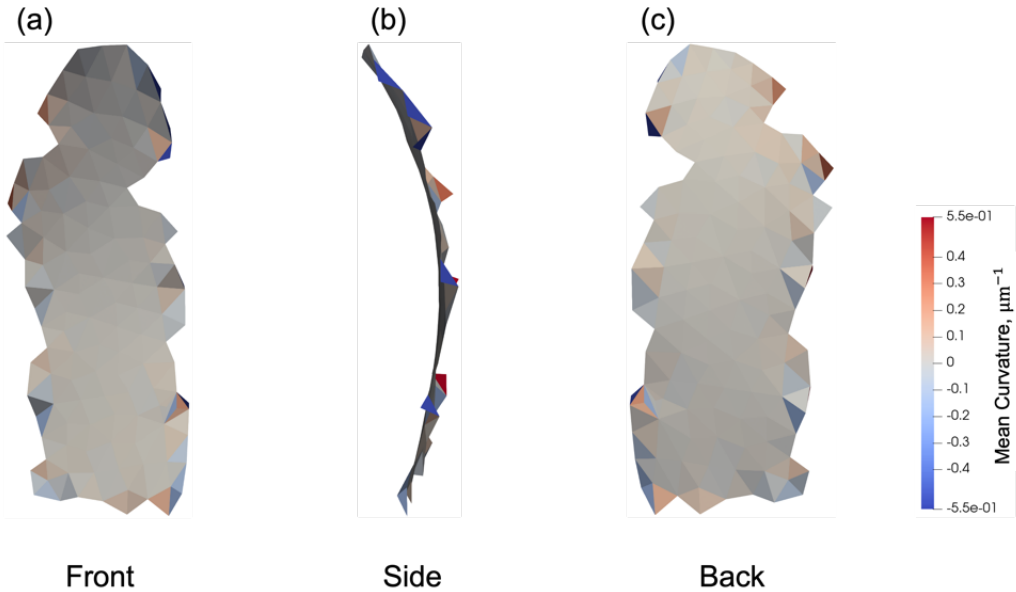

**Figure S7.** Three-dimensional views of the shape of the grain boundary in Fig. 10 after surface meshing. (a) Front view, (b) Side view and (c) Back view. Each triangle in the mesh is colored based on its mean curvature.

Fig. S7 illustrates the shape of the boundary discussed in Fig.10 after surface meshing. It clearly shows that, other than the triangles near triple junctions, the overall boundary curvature distribution is approximately uniform.